\documentclass[12pt]{article}
\def\be{\begin{equation}}
\def\ee{\end{equation}}
\def\la{\label}
\def\bea{\begin{eqnarray}}
\def\eea{\end{eqnarray}}
\def\non{\nonumber}
\def\ci{\cite}
\def\la{\label}
\def\bib{\bibitem}

\def\lm{\lambda}
\def\Lm{\Lambda}
\def\le{\left}
\def\ri{\right}

\def\S{\Sigma}
\def\gm{\gamma}

\def\al{\alpha}
\def\Omp{\Omega_\phi}
\def\Ompi{\Omega_{\phi i}}

\def\Om{\Omega}
\def\wp{w_\phi}

\def\Ompo{\Omega_{\phi o}}
\def\wpo{w_{\phi o}}
\def\weff{w_{eff}}

\def\fr{\frac}
\def\pp{\partial}
\def\raw{\rightarrow}

\usepackage{graphicx}

\begin{document}

\begin{flushright}
  hep-ph/0111292 \\

\end{flushright}

\vspace{15mm}

\begin{center}
   {\Large \bf   Quintessence Unification Models
    from Non-Abelian Gauge  Dynamics
   }

\end{center}

\vspace*{0.7cm}

\begin{center}
{\bf A. de la Macorra\footnote{e-mail: macorra@fisica.unam.mx}
}
\end{center}

\vspace*{0.1cm}

\begin{center}
\begin{tabular}{c}
{\small Instituto de F\'{\i}sica, UNAM}\\ {\small Apdo. Postal
20-364, 01000  M\'exico D.F., M\'exico}\\
\end{tabular}
\end{center}

\vspace{1 cm}

\begin{center}
{\bf ABSTRACT}
\end{center}
\small{We show that the condensates
of a non-abelian
gauge group, unified with the standard model gauge groups,
can parameterize the present day cosmological constant and
play the role of quintessence. The models agree with SN1a
and recent CMB analysis.

These
models have {\it no free} parameters.
Even the initial energy density  at the unification scale and at
the condensation scale are fixed by the number of degrees
of freedom of the gauge group (i.e. by $N_c, N_f$).
The values of $N_c, N_f$ are determined by imposing gauge
coupling unification and
 the number of models is quite
limited. Using Affleck-Dine-Seiberg superpotential
one obtains a scalar potential
$V=\Lm_c^{4+n}\phi^{-n}$.
Models with $2<n<4.27$ or equivalently $2\times 10^{-2}
GeV < \Lm_c < 6 \times 10^3 GeV$  do  not satisfy the unification
constrain. In fact, there are only three models and they have an
inverse power potential with $6/11 \leq n \leq 2/3$. Imposing
primordial nucleosynthesis bounds the preferred model has $N_c=3,
N_f=6$, with $n=2/3$,  a condensation scale $\Lm_c=4.2\times
10^{-8} GeV$ and $\wpo=-0.90$ with an average value $\weff=-0.93$.
Notice that the tracker solution is not a good approximation since
it has $w_{tr}=-\fr{2}{n+2}=-0.75$ for $n=2/3$.

We study the evolution of all fields from the unification
scale and we calculate the relevant cosmological quantities.
We also discuss the supersymmetry breaking mechanism which
is relevant for these models.

 }


\noindent \rule[.1in]{14.5cm}{.002in}

\thispagestyle{empty}

\setcounter{page}{0} \vfill\eject

\section{INTRODUCTION}

 The Maxima and Boomerang \ci{CMBR} observations on the cosmic
  microwave background radiation ("CMBR") and
the superonovae project SN1a \ci{SN1a}
have lead to conclude that the
universe is flat and it is expanding with an accelerating
velocity.  These conclusions show that the universe is now
dominated by a energy density with negative pressure with
$\Om_{\phi}=0.7 \pm 0.1$ and $ w_\phi < -2/3$ \ci{w}. New
analysis on the CMBR peaks constrain the models to have
$\wpo=-0.82^{+.14}_{-.11}$ \ci{neww}. This energy
is generically called the cosmological constant. Structure
formation also favors a non-vanishing cosmological constant
consistent with SN1a and CMBR observations \ci{structure}.
An interesting parameterization of this energy density is in
 terms of a scalar
field with gravitationally interaction only called quintessence
\ci{tracker}. The evolution of scalar field has been widely
studied and some general approach con be found in \ci{generic,mio.scalar}. The evolution of the scalar field $\phi$ depends
on the functional form of its potential $V(\phi)$ and a late time
accelerating universe constrains the form of the potential
\ci{mio.scalar}.

It is well known that the gauge coupling constant of a
non-abelian asymptotically free gauge group increases
with decreasing energy and the
free elementary fields will eventually
condense due to the strong interaction, e.g. mesons and baryons
in QCD. The scale where the coupling constant becomes strong is
called the condensation scale $\Lm_c$ and below it there are no more
free elementary fields. These condensates, e.g. "mesons",
develop a non trivial potential which can be calculated using
Affleck's potential \ci{Affleck}. The potential is of the form
$V=\Lm_c^{4+n}\phi^{-n}$, where $\phi$ represents the "mesons", and depending
on the value of $n$ the potential V may lead to an acceptable
phenomenology. The final value of $\wpo$ (from now on the subscript "o"
refers to present day quantities) depends $n$ and the initial
condition $\Ompi$ \ci{chris2}. A $\wpo < -2/3$, which is the upper limit of
 \ci{neww}, requires $n<2.74$ for $\Ompi \geq 0.25$ \ci{chris2}.
 For  smaller $\Ompi$ one obtains a larger $\wpo$ for a fixe $n$.
  The position
 of the third CMBR peak favors models with $n<1$ \ci{wette}
 and for some class of models
with $V=M^{4+n}\phi^{-n}e^{\phi^\beta/2}$, with $n\ge 1, \beta
\geq 0$, the constraint an the equation of state
 is even stricter $-1 \leq \wpo \leq -0.93$ \ci{wcop}.
In this kind of inverse power potential models (i.e. $n<2$)
the tracker solution is not a good approximation to the numerical
solution because the scalar field has not reached its tracker value
by present day.

Here we focus on a non-abelian asymptotically free gauge group whose
gauge coupling constant is unified with the couplings of the
standard model ("SM") ones \ci{chris1,chris2}.
 We will call this group the
quintessence or $Q$ group.
The cosmological picture in this
case is very pleasing. We assume gauge coupling unification
at the unification  scale $\Lm_{gut}$ for all gauge groups (as predicted by string
theory) and then let all fields evolve. At the beginning all
fields, SM and $Q$ model, are
massless and red shift as radiation until we reach the condensation
scale $\Lm_c$ of Q. Below this scale the fields of the quintessence
gauge group
will dynamically condense and we use Affleck's potential to study
its cosmological evolution. The
energy density of the $Q$ group $\Omp$ drops quickly,
independently of its initial conditions, and it is close to zero
for a long period of time, which  includes  nucleosynthesis (NS)
if $\Lm_c$ is larger than the NS energy $\Lm_{NS}$ (or temperature
$T_{NS}=0.1-10 MeV$), and becomes relevant only
until very recently. On the other hand, if $\Lm_c <  \Lm_{NS}$ than
the NS bounds on relativistic degrees of freedom must be imposed
on the models. Finally, the energy density of $Q$ grows and it
dominates at present time the total energy density with
the  $\Ompo \simeq 0.7$ and a negative pressure $\wpo <-2/3$
leading to an accelerating universe \ci{w}.

The initial conditions
at the unification scale and at the condensation scale
are  fixed by the number of degrees of freedom of the models
given in terms of $N_c, N_f$. Imposing gauge coupling unification
fixes $N_c, N_f$ and we do not have any free parameters in the
models (but for the susy breaking mechanism which we will
comment in section \ref{th}). It
is surprising that such a simple model works fine.

The restriction on $N_c, N_f$ by gauge unification rules out
models with a condensation energy scale between $2\times 10^{-2}
GeV < \Lm_c < 6 \times 10^3 GeV$ or for models with $2<n<4.27$
(the scale $\Lm_c$ is given in terms of $H_o$ and $n$ by
$\Lm_c\simeq H_o^{2/(4+n)}$ \ci{bine},\ci{chris2}). Since $\wpo
<-2/3$ requires $n<2.74$ all models must then have $\Lm_c <
2\times 10^{-2} GeV$. The number of models that satisfy gauge
coupling unification with a $\wpo <-2/3$ is quite limited and in
fact there are only three different models \ci{chris2}. All
acceptable models have $n\leq 2/3$ which implies that the
condensation scale is smaller than the NS scale. The preferred
model has $N_c=3, N_f=6$, $n=2/3$ and it gives $\wpo=-0.90$ with
an average value $\weff=-0.93$ agreeing with recent  CMBR analysis
\ci{neww,wette}.

It is worth mentioning that we have taken $\Lm_c$ as the one loop
renormalization energy scale (as used by Affleck et al
\ci{Affleck}) and if we had used the all loop renormalization
energy scale \ci{shifman} the values of $N_c, N_f$ of the models
may differ slightly but the general picture remains the same, i.e.
there are only a few models that satisfy the requirement of gauge
coupling unification, non of them have $n >2$ and there are no
free parameters.

\section{Condensation Scale and Scalar Potential}

We start be assuming that the universe has a matter content
of the supersymmetric gauge groups
$SU(1)\times SU(2)\times SU(3)\times SU(Q)$
where the first three are the SM gauge groups while the last one
corresponds to the "quintessence group" $Q$ and that the
couplings are unified at $\Lm_{gut }$ with $ g_1=g_2=g_3=g_Q=g_{gut}$.

The condensation scale $\Lm_c$ of a  gauge
group $SU(N_c)$ with $N_f$ (chiral + antichiral) matter fields
has in $N=1$ susy a one-loop renormalization group equation
given by
\be \la{lm}
\Lm_c= \Lm_{gut} e^{-\fr{8\pi^2}{b_o g^2_{gut}}}
 \ee
where $b_o=3 N_c-N_f$ is the one-loop beta function and
$\Lm_{gut}, g_{gut}$ are the unification energy scale and coupling
constant, respectively. From gauge coupling unification we know
that $\Lm_{gut} \simeq 10^{16}\mathrm{GeV}$ and $g_{gut} \simeq
\sqrt{4 \pi/ 25.7}$ \ci{unif}.

A phase transition takes place at the condensation scale
$\Lm_c$, since the elementary fields   are free fields above $\Lm_c$
and condense at $\Lm_c$. In order to study the cosmological
evolution of these condensates, which we will call $\phi$, we
use Affleck's potential \ci{Affleck}. This potential
is  non-perturbative and  exact \ci{duality}.

The superpotential for a non-abelian
$SU(N_c)$ gauge group with $N_f$ (chiral + antichiral) massless
matter fields  is  \ci{Affleck}
\be \la{super}
W=(N_c-N_f)(\fr{\Lm_c^{b_o}}{det <Q\tilde Q>})^{1/(N_c-N_f)}
\ee
where $b_o$ is the one-loop beta function coefficient. Taking $det
<Q\tilde Q>=\Pi_{j=1}^{N_f} \phi^2_j$ one has
$W=(N_c-N_f)(\Lm_c^{b_o}\phi^{-2N_f})^{1/(N_c-N_f)}$. The
 scalar potential in global supersymmetry is
$V=|W_\phi|^2$, with $W_\phi=\pp W/\pp \phi$, giving \ci{bine,mas}
\be
V=c^2\Lm_c^{4+n}\phi^{-n}
 \la{v} \ee
with $c=2N_f$, $n=2+4\fr{N_f}{N_c-N_f}$ and $\Lm_c$ is the
condensation scale of the gauge group $SU(N_c)$. The natural
initial value for the condensate is $\phi_i=\Lm_c$ since it is
precisely $\Lm_c$ the relevant scale of the physical process of
the field binding.

In eq.(\ref{v}) we have taken $\phi$ canonically normalized,
however the full Kahler potential $K$ is not known and for $\phi
\simeq 1$ other terms may become relevant \ci{bine} and could
spoil the runaway and quintessence behavior of $\phi$. Expanding
the Kahler potential as a series power $K=|\phi|^2+\Sigma_{i}a_i
|\phi|^{2i}/2i$ the canonically normalized field $\phi'$ can be
approximated\footnote{The canonically normalized field $\phi'$ is
defined as $\phi'=g(\phi,\bar{\phi})\phi$ with
$K_\phi^\phi=(g+\phi g _\phi+\bar{\phi}g_{\bar{\phi}})^2$} by
$\phi'=(K_\phi^\phi)^{1/2}\phi$ and  eq.(\ref{v}) would be given
by $V=(K_\phi^\phi)^{-1}|W_\phi|^2=(2 N_f)^2\Lm_c^{4+n}\phi^{-n}
(K_\phi^\phi)^{(n/2-1)}$. For $n<2$ the exponent term of
$K_\phi^\phi$ is negative so it would not spoil the runaway
behavior of $\phi$ \ci{chris1,chris2}.

If we wish to study models
 with $0<n<2$,
which are cosmologically favored \ci{chris2}  we need to consider
the possibility that not all $N_f$ condensates $\phi_i$ become
dynamical  but only a fraction $\nu$ are (with $N_f\geq\nu\geq 1$)
and we also need $N_f > N_c$ \ci{chris1,chris2}. It is important
to point out that even though it has been argued that for
$N_f>N_c$ there is no non-perturbative superpotential $W$
generated \ci{Affleck}, because the determinant of $Q\tilde Q$ in
eq.(\ref{super}) vanishes, this is not necessarily  the case
\ci{ax.asy}. If we consider the elementary quarks $Q_i^\al, \tilde
Q_i^\al$ ($i,j=1,2,...,N_f, \;\al=1,2,...,N_c$) to be the relevant
degrees of freedom, then for $N_c < N_f$ the quantity
$det(Q^i_\al\tilde Q^\al_j)$  vanishes since, being  the sum of
dyadics, always has zero eigenvalues. However,  we are interested
in studying the effective action for  the  "meson" fields
$\phi^i_j=<Q^i_\al \tilde Q^\al_j>$, and the determinant of
$\phi^i_j$, i.e. $det<Q^i_\al \tilde Q^\al_j>$, being the product
of expectation values does not need to vanish when $N_c <N_f$ (the
expectation of a product of operators is not equal to the product
of the expectations of each operator).

One can have $\nu \neq N_f$ with a gauge group with unmatching
number of chiral and anti-chiral fields or if some of the chiral
fields are also charged under another gauge group. In this  case
we have $c=2\nu, n=2+4\fr{\nu}{N_c-N_f}$ and $N_f-\nu$ condensates
fixed at their v.e.v. $<Q \tilde Q>=\Lambda_c^2$ \ci{chris1}.
Another possibility is by giving a mass term to $N_f-\nu$
condensates  $\varphi=<\bar Q_k Q_k>, \;(k=1,...,N_f-\nu$) while
leaving $\nu$ condensates $\phi^2=<\bar Q_j Q_j>,\;(j=1,...,\nu)$
massless. Notice that we have chosen a different parameterization
for $\varphi$ and $\phi$. The mass dimension for $\varphi$ is 2
while for $\phi$ it is 1. The superpotential now reads,
\be
\la{super2}
W=(N_c-N_f)(\fr{\Lm_c^{b_o}}{\phi^{2\nu}\varphi^{N_f-\nu}})^{1/(N_c-N_f)}+m\varphi
\ee
with $m$ the mass of $\bar Q_k Q_k$. If we take the natural choice
$\phi_i=\Lm_c$, as discussed above,  and $m=\Lm_c$ \ci{chris1} and
we integrate out the condensates $\varphi$ using
\be\la{varphi}
\fr{\pp W}{\pp \varphi}
=\varphi^{-1} \le((\nu-N_f)
\Lm_c^{(b_o-2\nu)/(N_c-N_f)}\varphi^{-(N_f-\nu)/(N_c-N_f)}+m\varphi\ri)=0
\ee
we obtain $\varphi=(N_f-\nu)^{(N_c-N_f)/(N_c-\nu)}\Lm_c^2$. By
integrating out the $\varphi$ field the second terms in
eq.(\ref{super2}), which is proportional to the first term, can be
eliminated. Substituting the solution of eq.(\ref{varphi}) into
eq.(\ref{super2}) one finds
\be\la{super3}
W=(N_c-\nu)(N_f-\nu)^{(N_f-\nu)/(N_c-\nu)}\Lm_c^{3+a}\phi^{-a}
 \ee
with $a=2\nu/(N_c-N_f)$.

The  scalar potential $V=|\pp W|^2$ is now given by
 \be\la{v2}
V=c'^2\Lm_c^{4+n'}\phi^{-n'}
 \ee
 with
$c'^2=4\nu^2(\fr{N_c-\nu}{N_c-N_f})^{2}(N_f-\nu)^{(N_f-\nu)/(N_c-\nu)}$
and $n'=2+4\nu/(N_c-N_f)$. Notice that for $\nu=N_f$ we recover
eq.(\ref{v}). From now on we will work with eq.(\ref{v2}) and we
will drop the quotation on $n'$.

The radiative corrections to the scalar potential eq.(\ref{v2})
are  $V\sim \Lm_c^{4+n}\phi^{-n}(1+O(\Lm_c^2\phi^{-2}))$
\ci{brax}. They are not important because we have $\phi \geq
\Lm_c$ and are negligible at late times when $\phi \gg \Lm_c$.

\subsection{Gauge Unification Condition}

 In order to have a model with gauge coupling
unification the scale $\Lm_c$ given in eq.(\ref{v}) or (\ref{v2})
must be identified with the energy scale in eq.(\ref{lm}).
However,  not all values of $\Lm_c$ will give an acceptable
cosmology. The correct values of $\Lm_c$ depend on the
cosmological evolution of the scalar condensate $\phi$ which is
determined by the power $n$ in eq.(\ref{v2}). The $\Lm_c$ scale
can be expressed in terms of present day quantities by \ci{chris2}
\be\la{lm2}
\Lm_c=\le(3Ho^2yo^2\phi_o^n c^{-2}\ri)^{\fr{1}{4+n}}
\ee
where $y^2$ is the fraction of the total energy density carried in
$V$, $y^2 \equiv V/3H^2=\Omp(1-\wp)/2$, and for
$\Ompo=0.7,\,\wpo=-2/3$ one has $yo=0.76$. A rough estimate of
eq.(\ref{lm2})  gives $\Lm_c\simeq H_o^{2/(4+n)}$ since we also
expect $\phi_o=O(1)$ today (we are living  at the beginning  of an
accelerating universe). The number of models that satisfy the
unification and cosmological constrains of having $\Ompo=0.7,
h_o=0.7$ (with the Hubble constant given by $H_o=100 h_o$ km/Mpc\,
sec) and $\wpo < -2/3$ \ci{w} is quite limited \ci{chris2}. In
fact there are only three models given in table \ref{constr}. These
models are obtained by equating  $\Lm_c$ from eq.(\ref{lm}), which
is a function of $N_c,\,N_f$ through $b_o$, and eq.(\ref{v2}),
which is also a function of $N_c,\,N_f, \nu$ through $n$. The
exact value of $y_o, \phi_o$ must be   determined by the
cosmological evolution of $\phi$ (c.f. eqs.(\ref{eqFRW})) starting
at $\Lm_c$ until present day.  For an acceptable model the
parameters $N_c, N_f$ and $\nu$ must take integer values.
We consider an acceptable model when $\Lm_c$ in eqs.(\ref{lm}) and
(\ref{lm2}) do not differ by more than 50\%. With this assumption
 there
are only 3 models, given in table \ref {constr}, that have
(almost) integer values for $N_f$. In all these models one has $n
\leq 2/3$ and the quantum corrections to the Kahler potential are,
therefore, not dangerous. All other combinations of $N_c, N_f,
\nu$ do not lead to an acceptable cosmological model.

From eq.(\ref{lm2}) one has for  $n\leq 4.27$ a scale $\Lm_c \leq
6.5\times 10^3 GeV$ and from eq.(\ref{lm}) this implies that
$b_o\leq   5.7$. Since $b_o=3N_c-N_f=2N_c+4\nu/(n-2)$ and the
minimum acceptable value for $N_c$ is two one finds $b_o \geq
4+4\nu/(n-2)$. Taking $2< n\leq 4.27$ gives a value of $b_o\geq
5.7$. The value of $n=4.27$ gives the upper limiting value for
which we can find a solution of eqs. (\ref{lm}) and (\ref{lm2}).
We see that it is not possible to have quintessence models with
gauge coupling unification with $2<n<4.27$. In terms of the
condensation scale the restriction for models with  $2\times
10^{-2} GeV < \Lm_c < 6 \times 10^3 GeV$.

Using $n=2+4\nu/(N_c-N_f)$ or equivalently $N_f=N_c+4\nu/(n-2)$
with $b_o=3N_c-N_f=2N_c+4\nu/(n-2)$ we can write from
eq.(\ref{lm}) as
$b_o=8\pi^2/g_{gut}^2(Log(\fr{\Lm_{gut}}{\Lm_c}))^{-1}$ and $N_c$
\bea\la{nc}
N_c&=&\fr{1}{2} b_o+\fr{2\nu}{2-n}\non\\
&=&\fr{4\pi^2}{g_{gut}^2} (Log[\fr{\Lm_{gut}}{\Lm_c}])^{-1}+\fr{2\nu}{2-n}
\eea
Form eq.(\ref{lm2}) we have $\Lm_c$ as a function of $n$ (with the
approximation of  $y^2_o\phi_o^n=1$) and $N_c$ in eq.(\ref{nc})
becomes a function of $n$ and  $\nu$ only. In figure \ref{fig2} we
show $N_c$ as a function of $n$ or  $\Lm_c$ with the constraint of
gauge coupling unification. We see that for  $2\times 10^{-2} GeV
< \Lm_c < 6.5\times 10^3 GeV$ we have a $N_c < 2$ and therefore
are ruled out. In terms of $n$ the condition is that models with
$2<n<4.27$ are not viable. In deriving these conditions, we have
taken $\nu=1$ which gives the smallest constraint to $N_c$ as seen
from eq.(\ref{nc}).

The upper limit $\Lm_c > 6.5\times 10^3 GeV$ has $n > 4.27$ (c.f.
eq.(\ref{lm2})). As mentioned in the introduction, the value of
$\wpo$ depends on the initial condition $\Ompi$ and on $n$
\ci{chris2}. The larger $n$ the larger $\wpo$ will be (same is
true for the tracker value $w_{tr}=-2/(2+n)$). It has been shown
that assuming an initial value of $\Ompi$ no smaller than 0.25
then the value of $\wpo$ will be less then $\wpo<-2/3$ only if $n
< 2.74$ \ci{chris2}. Therefore, the models with $n>4.27$ are not
phenomenological acceptable and since $4.27>n>2$ are also ruled
out by the constrain on gauge coupling unification, we are left
with models with
\be\la{lmn}
\Lm_c < 2\times 10^{-2} GeV  \hspace{1cm} or\hspace{1cm} n<2.
\ee
 So, only models with a
cosmological late time phase transition are allowed.

\begin{figure}[htp!]
\begin{center}
\includegraphics[width=6cm]{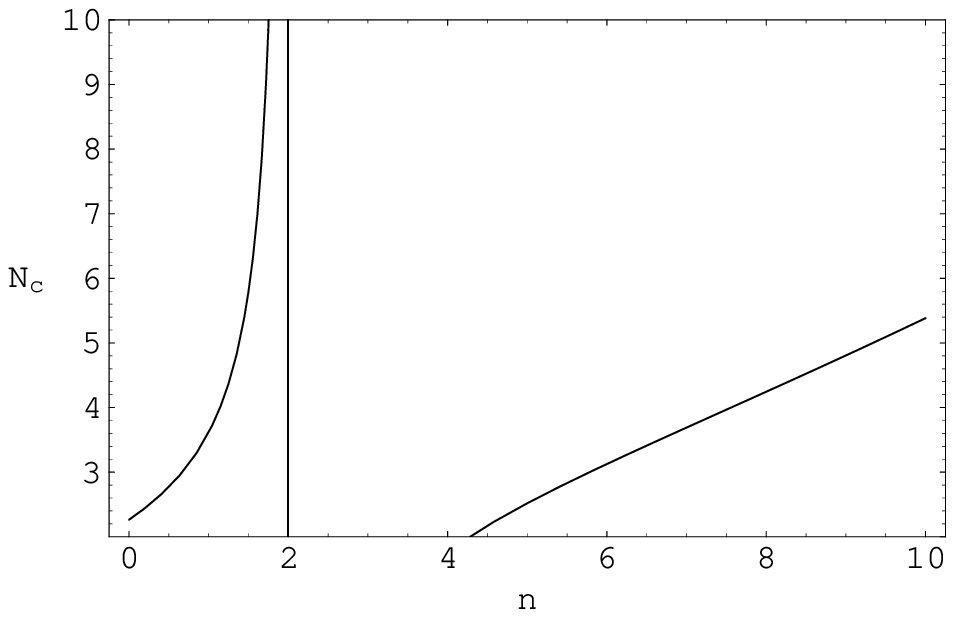}
\includegraphics[width=6cm]{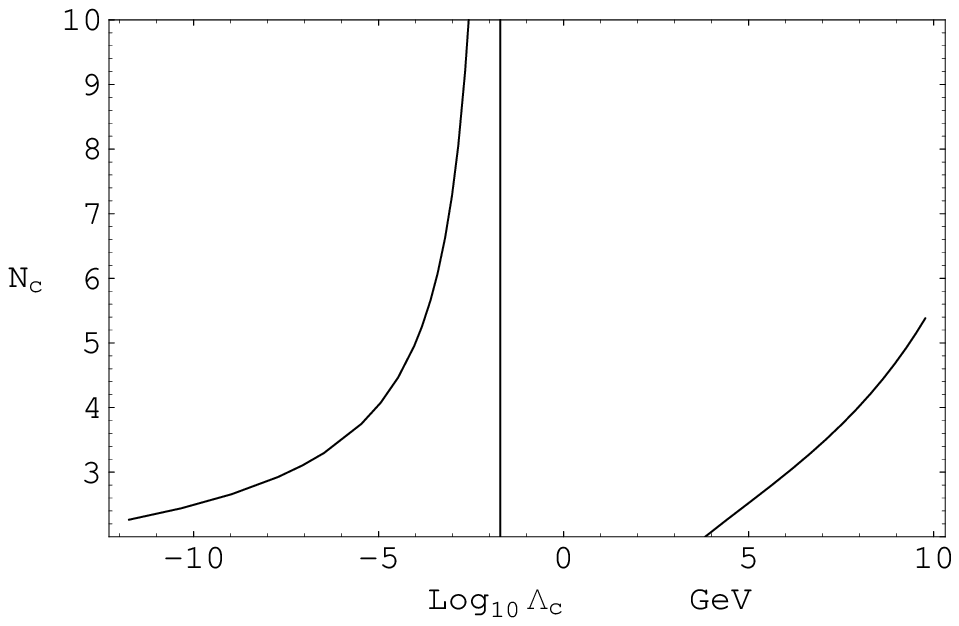}
\end{center}
\caption{\small{We show  $N_c$
as a function of $n$ and the energy scale $\Lm_c$  after
imposing gauge coupling unification. $N_c$ must
be larger than 2 and we have taken $\nu=1$.
}}
\la{fig2}
\end{figure}

\begin{table}\la{constr}
\begin{center}
\begin{tabular}{|c|c|c|c|c|}
\hline
 Num& $ N_c$ &$N_f$ &$ \nu$ & $n$ \\ \hline
\hline
 I& 3 & 5.98 & 1 & 0.66   \\
 II &6 & 14.97 & 3 & 0.66   \\
  III& 7 & 18.05 & 4 & 0.55  \\
  \hline
\end{tabular}
\end{center}
\caption{\small{Models that satisfy gauge coupling unification
and have $n< 2.74 $ (i.e.$\wpo <-2/3$)}}
\end{table}

\section{ Thermodynamics,
Nucleosynthesis Bounds and  Initial Conditions}\la{th}

Before determining the evolution of $\phi$ we must
analyze  the initial  conditions for the $SU(Q)$ gauge
group. The general picture
is the following: The $Q$ gauge group is by hypothesis, unified with
the SM gauge groups at the unification energy $\Lm_{gut}$. For energies
scales between the unification and condensation scale, i.e.
$\Lm_c <\Lm <\Lm_{gut}$, the elementary fields of $SU(Q)$
are massless and weakly coupled and interact with the SM only gravitationally.
The Q gauge interaction becomes
strong at $\Lm_c$ and condense the elementary fields leading
to the potential in eq.(\ref{v2}).

Since for energies above $\Lm_{gut}$ we have a single gauge group
it is naturally to assume that all fields (SM and Q) are in
thermal equilibrium. However,  at temperatures $T < T_{gut}$ the
gauge group $Q$ is decoupled since it interacts with the SM only
via gravity.

The energy density at  the unification scale is
given by $\rho_{Tot}=\fr{\pi^2}{30}g_{Tot}T^4$, where
$g_{Tot}=\S Bosons+7/8 \S Fermions$ is the
total number of degrees of freedom at the temperature $T$.
The minimal models have $g_{Tot}=g_{sm  i}+g_{Q i}$, with
$g_{sm  i}=228.75$ and $g_{Q i}=(1+7/8)(2(N_c^2-1)+2N_fN_c)$
for the minimal supersymmetric standard model MSSM and for the
$SU(Q)$ supersymmetric gauge group with $N_c$ colors and $N_f$ (chiral
+ antichiral) massless fields, respectively.
The initial energy density at the unification scale  for each group
is simply given in terms of number of degrees of freedom,
$\Om=\rho/\rho_c$,
\be\la{in.con}
\Om_{Q i}(\Lm_{gut})=\fr{g_{Q i}}{g_{Tot}},\space{.5cm}
\;\;\Om_{sm i}(\Lm_{gut})=\fr{g_{sm  i}}{g_{Tot}}
\ee
with $\Om=\Om_{Q}+\Om_{sm}=1$. Since the SM and $Q$ gauge groups
are decoupled below $\Lm_{gut}$, their respective entropy,
$S_k=g_ka^3T^3$ with $g_k$ the degrees of freedom of the $k$ group
and $a$ the scale factor of the universe (see eq.(\ref{eqFRW2})),
will  be independently conserved. The total energy density $\rho$
as a function of the  photon's temperature $T$ above $\Lm_c$ (i.e.
$\Lm_c < \Lm < \Lm_{gut}$),  with the $Q$ fields still massless
and redshifting as radiation, is given by
\be\la{g*}
\rho=\fr{\pi^2}{30}g_{*}T^4
\ee
with
\be
g_*=g_{sm f}+ g_{Q f} \le(\fr{T'}{T}\ri)^4
=g_{sm f}+ g_{Q f}\le(\fr{g_{sm f}g_{Q i}}{g_{sm i}g_{Q f}}\ri)^{4/3}
\ee
and  $g_{sm i}, g_{sm f}, g_{Q i}, g_{Q f}$ are the initial (i.e.
at the unification scale) and final standard model and $Q$ model
relativistic degrees of freedom, respectively. From the entropy
conservation, we know that the relative temperature between the
standard model and the $Q$ model is given by $T'/T=(\fr{g_{sm
f}g_{Q i}}{g_{sm i}g_{Q f}})^{1/3}$. It is clear that the energy
density for the $Q$ model $\rho_Q=\pi^2/30 g_Q T'^4$ in terms of
the photon's temperature $T$ is fixed
 by the number of degrees of freedom,
\bea\la{omq}
\Om_{Q f}&=&\fr{g_{Q f}T'^4}{g_*T^4}\non\\
&=& \fr{g_{Q f}(g_{sm f}g_{Q i}/g_{sm  i}g_{Q f})^{4/3}}
{g_{sm  f}+g_{Q f}(g_{sm f}g_{Q i}/g_{sm  i}g_{Q f})^{4/3}} .
\eea
Eq.(\ref{omq}) permits us to determine the energy density of the
$Q$ group at any temperature above the condensation scale.

\subsection{Energy Density at $\Lm_c$}

We would like now to determine the energy density at the
condensation scale which will set the initial energy density for
the scalar composite field $\phi$.

Just above the condensation scale $\Lm_c$ we take, for simplicity
of argument, that all particles in the $Q$ group are still
massless and we can use eq.(\ref{omq}) to determine the
$\Om_Q(\Lm_c)$ with $g_{Q i}=g_{Q f}$ giving
\be\la{omq2}
\Om_{Q f} =\fr{g_{Q f}(g_{sm f}/g_{sm  i})^{4/3}} {g_{sm f}+g_{Q
f}(g_{sm f}/g_{sm  i})^{4/3}}.
\ee
At $\Lm_c$ we have a phase transition and we no longer have
elementary free particles in the $Q$ group. They are bind together
through the strong gauge interaction and the acquire a
non-perturbative potential and mass given by eq.(\ref{v2}). In
other words, below the condensation scale there are no free
"quarks" $Q$ and we have "meson" and "baryon" fields.

If we consider only  the SM and the Q group, the energy density
within the particles of the Q group must be conserved since they
are decoupled from the SM (the interaction is by hypothesis only
gravitational).

Furthermore,  the "baryons", which  we expect to be heavier then
the lightest meson field (as in QCD), and the massive "meson"
fields $\varphi$ (see eq.(\ref{super2})) are coupled (i.e.
$\Gamma/H >1$) to the lightest composite field $\phi$ for
temperatures  $\Lm_{gut} > T > \Lm_c(\Lm_c/m_p)^{1/3}$, with
$\Lm_c/m_p \ll 1$. The "baryons" and the heavy $\varphi$ fields
will then  decay into the lightest state within the $Q$ group,
i.e. the  $\phi$ field. So, we conclude that all the energy of the
$Q$ group is transmitted into $\phi$ at around the phase
transition scale given by  condensation scale $\Lm_c$ and
\be
\Om_\phi(\Lm_c)=\Om_Q(\Lm_c).
\ee
This is a natural assumption from a particle point of view but is
not crucial from a cosmological point of view, in the sense that
any "reasonable" fraction of the energy density in the $Q$ group
would give a correct cosmological evolution of the $\phi$ field.

We would like to stress out that the initial condition for $\phi$
is no longer a  free parameter but it is given in terms of the
degrees of freedom of the MSSM and the $Q$ group.

\subsection{ Nucleosyntehsis Constrain on $\Om_Q$}

The big-bang nucleosynthesis (NS) bound on the energy density from
non SM fields, relativistic or  non-relativistic, is quite
stringent $\Om_{Q} < 0.1$ \ci{NS} and a recent more conservative
bound gives $\Om_Q < 0.2$ \ci{NSC}.

If the $Q$ gauge group condense at temperatures much higher than
NS then, the evolution of the condensates will be given by
eqs.(\ref{eqFRW2}) with the potential of eq.(\ref{v}) and we must
check that $\Om_Q$ at NS is no larger than 0.1-0.2. This will be,
in general,  no problem since it was shown that even for a large
initial $\Om_Q$ at the condensation scale the evolution of $\phi$
is such that $\Om_Q$ decreases quite rapidly and remains small for
a long period of time (see figure \ref{fig1}) \ci{chris1,chris2}.

On the other hand, if the gauge group condenses after NS we must
determine if the $Q$ energy density is smaller than $\Om_Q <
0.1-0.2 $ at NS. Since the condensations scale $\Lm_c$ is smaller
than the NS scale, all fields in the $Q$ group are still massless
and the energy density is given in terms of the relativistic
degrees of freedom  and from  eq.(\ref{omq}) to set a limit on $
g_{Q f}$ and $g_{Q i}$,
\be \la{dg}
\Delta g_Q \equiv g_{Q f}^{-1/3}g_{Q i}^{4/3} =
\fr{\Om_Q}{1-\Om_Q}g_{sm f}^{-1/3}g_{sm  i}^{4/3}
\ee
and for $g_{Q f}=g_{Q i}=g_Q$
\be \la{dg2}
\Delta g_Q = g_{Q }= \fr{\Om_Q}{1-\Om_Q}g_{sm f}^{-1/3}g_{sm
i}^{4/3}
\ee
where we should take $g_{sm f}=10.75$ at the final stage (i.e. NS
scale) and $g_{sm  i}=228.75$ at the initial stage (i.e. at
unification)  for the minimal supersymmetric standard model MSSM.
For $\Om_Q\leq 0.1, 0.2$ eq.(\ref{dg}) gives un upper limit on the
number of relativistic degrees of freedom $\Delta g_Q \leq 70,
158$ respectively (or $g_{Q } \leq 70, 158$ if  $g_{Q f}=g_{Q
i}=g_Q$).

The l.h.s. of eq.(\ref{dg}) depends on the initial (i.e. at unification)
and final (at NS) number of degrees of freedom of the gauge group Q.
The smaller (larger) the initial (final) degrees of freedom of $Q$
the smaller $\Delta g_Q$ and $\Om_Q$ will be.

\subsection{Supersymmetry Breaking}\la{susy}

 Another
important ingredient in these models is the way supersymmetry is
broken. The precise mechanism for susy breaking is still an open
issue but it is generally believed that  gaugino condensation of a
non-abelian gauge group breaks susy \ci{gaugino}.  There are a two
ways that the breaking of susy is transmitted to the MSSM, by
gravity \ci{sus.grav} or via gauge interaction \ci{sus.gauge}.

In the case of gravity susy breaking, the same mechanism that
breaks susy for the MSSM will break susy for the $Q$ group and
from particle physics we expect the breaking to be transmitted at
$m\sim\Lm^3_{break}/m_p^2 \sim TeV$  scale (i.e.
$\Lm_{break}\simeq 10^{11}GeV$).
 The final degrees of freedom of the $Q$ group must contain only
 the non-supersymmetric ones at
temperatures $T < TeV$, with $g_{Q f}=2(N_c^2-1)+2 N_fN_c 7/8$ at
NS and the initial ones at unification are $g_{Q
i}=(1+7/8)(2(N_c^2-1)+2 N_fN_c)$. The Q group would be globally
supersymmetric but would have explicit soft supersymmetry breaking
terms (as the breaking of MSSM to SM). The fields in the gauge
group responsible for susy breaking are not in thermal equilibrium
at $T < T_{gut}$ neither with the SM nor with the Q group since
they interact via gravity only.

On the other hand if susy breaking is  gauge mediated and since
the $Q$ group interacts only gravitationally with all other gauge
groups, the  supersymmetry  breaking for $Q$ group will be at a
scale $m\sim\Lm^3_{break}/m_p^2 \sim 10^{-15}GeV$, since one
expects the condensation scale of the susy breaking gauge group to
be in this case much smaller than for the gravity one, with
$\Lm_{break}\leq O(10^7 GeV)$ \ci{sus.gauge}, to give a susy
breaking mass to the SM of the order of TeV. Therefore, in this
second case the $Q$ group will be supersymmetric for models with
$\Lm_c > m \sim 10^{-15} GeV$ and the relativistic degrees of
freedom at NS will be the same as the initial ones, i.e. $g_{Q
f}=g_{Q i}=(1+7/8)(2(N_c^2-1)+2 N_fN_c)$ at NS. If susy breaking
is gauge mediated, then the gauge group responsible for susy
breaking will be coupled to the MSSM and will be in thermal
equilibrium at $\Lm_c< T \leq T_{gut}$ and its degrees of freedom must be
taken into account in the initial $g_{sm i}=g_{SM i} + g_{ex}$,
where $g_{SM i}$ are the degrees of freedom of the MSSM and $g_{ex}$
those of the gauge group responsible for susy breaking. Typical
models of susy breaking via gauge interaction have a gauge group
$SU(N_c)$ with $N_c>5$ and $N_f>4$ \ci{sus.gauge} which gives and extra
$g_{ex}\geq 160$.

We would like to point out that in both cases the susy breaking
mass is a problem for quintessence since the present day mass must
be  of the order of $10^{-33} GeV$, much smaller than the susy
breaking mass. Here, we have nothing new to say about this problem
and we consider it as part of the ultraviolet cosmological
constant problem, i.e. the stability of the vacuum  energy
(quintessence energy) to all quantum corrections. The contribution
to the scalar potential from the susy breaking scale from the
$\phi$ field and/or  from any other field of the MSSM is enormous
compared to the required present day value. The ultraviolet
problem is an unsolved and probably one the most important
problems in theoretical physics.

\subsection{Models}

 Now, let us determine the contribution to the energy density at
NS for the three models given in table \ref{tab1}, taking
the closest integer for $N_f$. The number of degrees of freedom
for an $SU(N_c)$ supersymmetric gauge group with $N_f$ flavors is
$g_{Q i}=(1+7/8)(2(N_c^2-1)+2N_fN_c)$. All three
models have the same supersymmetric one-loop beta function
$b_o=3N_c-N_f=3$ and $\Lm_c=4\times 10^{-8} GeV$.

The group with the smallest number of degrees of freedom is Model
I, $N_c=3, N_f=6$ and we have $g_{Q s}=97.5$
 supersymmetric   degrees of
freedom.  In this model, we have  at NS $\;  \Om_{Q} |_{NS}=0.13$. We
see that  the energy density of $Q$ is slightly larger than the
stringent NS bound $\Om_{Q} |_{NS} < 0.1$ but it is ok with the
more conservative bound $\Om_Q<0.2$.

For other thwo groups we have $\Om_{Q} |_{NS}=0.42,0.51$ for
Models II and  III  respectively. $\Om_Q|_{NS}$ is larger since they
have a larger $g_{Q i}$ (i.e. $N_c, N_f$) and all these models would
not satisfy the NS energy bound $\Om_{Q} |_{NS}<0.2$. Therefore,
if susy is broken via gravity these three models would not be
phenomenologically viable, unless more structure is included. On
the other hand, if susy is broken via gauge interaction we would need
to take into account  the
degrees of freedom of the susy breaking group given in $g_{sm
i}=g_{SM i}+g_{ex}$ when calculating $\Om_{Q} |_{NS}$.
These extra degrees of freedom give a larger
$g_{sm i}$ and therefore reduce $\Ompi$ as can be seen from
eq.(\ref{omq}). In order to have $\Om_{Q} |_{NS}\leq 0.1$ we
require $g_{ex}\geq 64, 718,986$ for models I, II, and
III, respectively, while for  $\Om_{Q} |_{NS}\leq 0.2$ we
require $g_{ex}\geq 287, 433$ for models II and
III, respectively.

We have checked that a larger number of extra degrees of freedom
does not affect the cosmological evolution of $\phi$
significantly.  In fact there is no "reasonable" upper  limit on
$g_{ex}$ from the cosmological point of view (e.g. for
$g_{ex}=10^9$ the model is still ok) as can be seen in
fig.\ref{fig1}. Notice that a large $g_{ex}\gg 1$ gives a small
energy density $\Ompi \propto g_{sm i}^{-4/3} \propto
g_{ex}^{-4/3}$.  This result also shows that an acceptable
cosmological model cosmological is almost independent
on the initial energy density of $\phi$.

As a matter of completeness we give the minimal model when susy is
broken via gravity. In this case one has to consider that $g_{Q
f}\neq g_{Q i}$ as discussed in section \ref{susy}. The minimal
gauge group when susy is broken via gravity has $N_c=5, N_f=14,\nu=1$
and $g_{Q s}=352.5, \;g_{Q ns}= 170.5$ for the relativistic susy
and non-susy degrees of freedom, respectively, and one has $n=14/9
\simeq 1.5,\; \Lm_c=4.5 \times 10^{-4} GeV,\; \Om_Q |_{NS}=0.41$
much larger than the NS bound. The difference in the values of
$N_c, N_f$ between the susy and non-susy models are due to a
change in $b_o$, the one loop-beta function in eq.(\ref{lm}),
below the susy breaking scale $1 TeV$, giving different values for
$\Lm_c$ for the same $N_c, N_f$. We conclude that unless more
structure is included (i.e. need $g_{ex}=689$ relativistic fields
coupled to the SM to have $\Om_Q|_{NS} <0.1$) there are no models
that satisfy the NS energy bound  for the case when  susy is
broken via gravity. However, if we allow for a discrepancy
in $\Lm_c$ from eqs.(\ref{lm}) and (\ref{lm2}) of
up to one order of magnitude then the model $N_c=3, N_f=9, \nu=3$
would be fine and it has  $g_{Q s}=131.25,
 \;g_{Q ns}= 63.25$ for the relativistic susy
and non-susy degrees of freedom, respectively, with  $n=4/3
,\; \Lm_c=1-9 \times 10^{-12} GeV,\; \Om_Q |_{NS}=0.2$.

\section{Cosmological Evolution of $\phi$}

The cosmological evolution of $\phi$ with an arbitrary potential
$V(\phi)$ can be determined from a system of differential
equations describing a spatially flat Friedmann--Robertson--Walker
universe in the presence of a barotropic fluid energy density
$\rho_{\gm}$ that can be either radiation or matter, are
\bea\la{eqFRW}
\dot H &=& -\frac{1}{2}(\rho_\gamma+ p_\gamma+\dot \phi^2),\non \\
\dot \rho &=& -3H(\rho+p),\\
\ddot \phi &=& -3H \dot \phi-\frac{d V(\phi)}{d \phi},\non
 \eea
where $H$ is the Hubble parameter, $\dot \phi = d\phi/d t$, $\rho$
($p$) is the total energy density (pressure).  We use
the change of variables $x \equiv \frac{\dot
\phi}{\sqrt{6} H}$ and $y \equiv \frac{\sqrt{V}}{\sqrt{3} H }$
   and  equations (\ref{eqFRW}) take the following form
  \cite{liddle,mio.scalar}:
 \bea  \la{eqFRW2}
x_N&=& -3 x +\sqrt {3 \over 2} \lambda\,  y^2 + {3 \over 2} x
[2x^2 + \gm_{\gm} (1 - x^2- y^2)] \non\\
 y_N&=& - \sqrt {3 \over
2} \lambda \, x\, y + {3 \over 2} y [2x^2 + \gm_{\gm} (1 - x^2
-y^2)]\\
H_N&=& -{3 \over 2} H [2x^2 + \gm_{\gm} (1 - x^2 - y^2)]
\non
 \eea
where $N$ is the logarithm of the scale factor $a$, $N\equiv
ln(a)$; $f_N\equiv d f/d N$ for $f=x,y,H$; $\gm_\gm= 1+ w_\gm $
and $\lambda(N)\equiv -V'/V$ with $V'= dV/d\phi$. In terms of $x,
y$ the energy density parameter is $\Om_{\phi}=x^2+y^2$ while the
equation of state parameter is given by $w_{\phi}\equiv
p_{\phi}/\rho_{\phi}=\frac{x^2-y^2}{x^2+y^2}$ (with $m_p^2=G/8\pi=1$).

The Friedmann or constraint equation for a flat universe
$\Omega_{\gm}+\Om_{\phi}=1$ must supplement equations
(\ref{eqFRW2}) which are valid for any scalar potential as long as
the interaction between the scalar field and matter or radiation
is gravitational only. This set of differential equations is
non-linear and for most cases has no analytical solutions. A
general analysis for arbitrary potentials is performed in
\cite{mio.scalar}, the conclusion there is that all model
dependence falls on two quantities: $\lambda(N)$ and the constant
parameter $\gm_\gm$. In the particular case given by $V \propto
1/\phi^n$ we find $\lm \raw 0$ in the asymptotic limit. If we
think the scalar field appears well after Planck scale we have
$\lm_i=n\,m_{Pl}/\phi_i=n\,m_{Pl}/\Lm_c \gg 1$  (the subscript $i$
corresponds to the initial value of a quantity). An interesting
general property of these models is the presence of a many e-folds
scaling period in which $\lm$ is practically a constant and
$\Om_{\phi} \ll 1$. After a long permanence of this parameter at a
constant value it evolves to zero, $\lm \raw 0$, which implies
$\frac{x_N}{x}<0$ and $\frac{y_N}{y}>0$ \cite{mio.scalar}, leaving
us with $\Om_{\phi} \equiv x^2+y^2 \raw 1$ and $\wpo \equiv
\frac{x^2-y^2}{x^2+y^2} \raw -1$, which are in accordance with a
universe dominated by a quintessence field whose equation of state
parameter agrees with positively accelerated expansion.

The evolution of $\Omp$  can be observed in
Figure \ref{fig1}, together with the evolution of $\wp$ which
fulfills the condition $\wpo < -2/3$ \ci{w} for different initial
conditions.

The value of the condensation scale in terms of $H_o$ is
\be \la{Lm}
\Lm_c=\left(\fr{3 y_o^2 H_o^2}{4\nu^2} \right)^\frac{1}{4+n}
\phi_{o}^\frac{n}{4+n}
 \ee
and together with eq.(\ref{lm}) sets a constrain for $N_c, N_f$.
The approximated value for $y_o^2,\phi_o$ can be obtained from
eq.(\ref{po}) but one expects in general to have $0.76< y_o< 0.83$
and $\phi_o\sim 1$ for $\Ompo=0.7$ and $\wpo<-2/3$. This can be
also seen from the identity $y^2= \Omp(1-\wp)/2$. The order of
magnitude   of the condensation scale is therefore
$\Lm_c=H_o^{2/(4+n)}$.

The value of $\wpo$ can be approximated by \ci{chris2}
 \be\la{wo}
 \wpo=-1+\fr{n^2 \Ompo}{3 \phi_o^2}
 \ee
 with $\phi_o$ given by solving \ci{chris2}
\be  \la{po}
  \phi_o^2 -\phi_{sc}^n\phi_o^{2-n}-\fr{n^2}{6}\Ompo=0
\ee
where $\phi_{sc}$ is the scaling value of $\phi$, i.e. the constant
value at which $\phi$ stays for a long period of time. The scaling
value is given only in terms of $\Ompi$, $
 \phi_{sc}=\phi_i+\sqrt{6\Ompi}$
  for $\Ompi < 1/2$ and
 $ \phi_{sc}=\phi_i + \sqrt{6}\le(\fr{1}{\sqrt{2}}+\fr{1}{2}Log[\fr{\Ompi}{1-\Ompi}]\ri)
$ for $\Ompi > 1/2$ \ci{tracker}.

In order to analytically solve eqs.(\ref{po}) we need to fix the
value of $n$ and we can determine $\wpo$ by putting  the solution
of (\ref{po}) into eq.(\ref{wo}). Eq.(\ref{po}) can be rewritten as
$\phi_o=\phi_{sc}(1-n^2\Ompo/6\phi_o^2)^{-1/n}$
and we see that $\phi_o > \phi_{sc}$. For
$\gm_{\phi o}=n^2\Ompo/6\phi_o^2 \ll 1$ one
has $\phi_o \simeq \phi_{sc}$  and for the simple cases of $n=1,2$ and $4$
we find $\phi_o|_{n=1}=\phi_{sc}/2+\sqrt{9\phi_{sc}^2+6\Ompo}/6,
\;y^2_o|_{n=1}=\phi_{sc}(-3\phi_{sc}+\sqrt{9\phi_{sc}^2+6\Ompo},\;)$
$\phi_o|_{n=2}=\sqrt{\phi_{sc}^2+2\Ompo/3},\;y^2_o|_{n=2}=
3\phi_{sc}^2\Ompo/(3\phi_{sc}^2+2\Ompo)$
and $\phi_o|_{n=4}=\sqrt{4\Ompo/3
+\sqrt{9\phi_{sc}^2+16\Ompo}/3},\;$
$y^2_o|_{n=4}=\Ompo-8\Ompo^2/(4\Ompo+\sqrt{9\phi_{sc}^2+16\Ompo})$,
respectively. Notice that the value of
$\phi_o, \wpo$ at $\Ompo=0.7$ {\it does not} depend on $H_i$ or
$H_o$ and it only depends on $\Ompi$ (through $\phi_{sc}$) and
$n$.

\section{The Models}

\begin{figure}[htp!]
\begin{center}
\includegraphics[width=8cm]{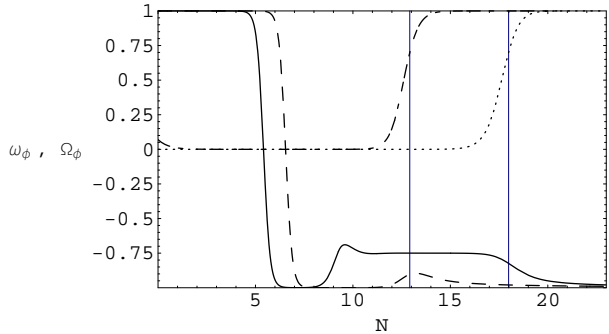}
\end{center}
\caption{\small{We show the evolution for $\Omp, \wp$
for initial condition $\Ompi=0.07$ dashed-dotted and
dashed lines, respectively and for $\Ompi=10^{-10}$, dotted
and solid lines, respectively, with $n=2/3$. The first case corresponds
to $g_{ex}=64$ while the later case has a huge number
of extra degrees of freedom $g_{ex}>10^9$. The vertical
lines correspond to present day values with $\Ompo=0.7$
and $h_o=0.7$.
}}
\la{fig1}
\end{figure}

In this section we study the three different models given in table
\ref{constr}. It is interesting to note that all three models have
a one-loop beta function coefficient $b_o=3N_c-N_f=3$ which
implies that they have the same  condensations scale $\Lm_c=4.2
\times 10^{-8} GeV$. The power of the exponent $n$, see table
\ref{constr}, is very similar and if we take the closest integer
value for $N_f$ one has $n=2/3$ or $ 6/11$. Notice that
model I is self dual $\tilde N_c=N_f-N_c=3$ with $N_f$ matter
fields. The other two models are not self dual.

From now on we will focus on the Model I of table \ref{tab1}
and we will summarize the relevant quantities in table
\ref{tab2} for all models.

The initial energy density at the unification scale is given by
eq.(\ref{omq}) with $g_{Qi}=97.5, g_{sm i}=228.75$ is
$\Om_Q(\Lm_{gut})=0.3$. Below $\Lm_{gut}$ the fields are weakly
coupled, massless (they redshift as radiation) and are decoupled
from the SM.
  A phase transition takes place when the gauge coupling constant
becomes strong at the condensation scale.
Since the condensation scale is much smaller than the NS scale,
$\Lm_c \ll 0.1 MeV$, we expect all fields of the  $Q$ group to be
relativistic at NS. From eq.(\ref{dg}) with $g_{sm f}=10.75$
 the energy density,
 assuming no extra degrees
of freedom, is $\Om_Q|_{NS}=0.13$ for susy Model I.
In order to satisfy the NS bound $\Om_Q|_{NS}< 0.1$, 64 extra relativistic degrees
of freedom in thermal equilibrium with the SM at $T \leq T_{gut}$
are required while for $\Om_Q|_{NS} <0.2$  the model does not require any
extra degrees of freedom.

What is the energy density of $Q$ at the condensation scale
$\Lm_c$? Using eq.(\ref{omq}) with $g_{sm f}=3.36, g_{Q f}=97.5$
given at $\Lm_c=4.2 \times 10^{-8} GeV $ and with no extra degrees
of freedom at the unification scale (i.e. $g_{sm i}=228.75$ and
$g_{Q i}=g_{Q f}=97.5$)  for Model I one has
$\Omp(\Lm_c)=\Om_Q(\Lm_c)=0.095$.  Imposing the stringent NS bound
$\Om_Q|_{NS} <0.1$ we need to include  $g_{ex}=64$ extra  degrees
of freedom (that should come from susy breaking mechanism) and the
energy at $\Lm_c$ is now  $\Omp(\Lm_c)=0.07$.

Evolving eqs.(\ref{eqFRW2}) with initial condition
$\Omp(\Lm_c)=0.07$ gives at present time with $h_o=0.7,\;
\Ompo=0.7$ a value of  $\weff \equiv\int da\
\Om(a)\wp(a)/\int da\ \Om (a)=-0.93$ with
of $\wpo=-0.90$  in agreement with SN1a and CMBR data. The
analytic solution given in eq.(\ref{wo}) is $\wpo(Th)=-0.82$ and
it is a much better approximation to the numerical value then
the  tracker value $w_{tr}=-2/(2+n)=-0.75$ \ci{tracker} which
is the upper
value of $\wpo$ for given $n$ and arbitrary initial conditions.

From a cosmological evolution point of view, we have a large range
of initial condition of $\Ompi$ \ci{chris2}. The upper limit is
set by NS and there is no "reasonable" lower limit (a smaller
$\Ompi$ implies that we have a much larger number of extra degrees
of freedom $g_{ex}$ but it must be finite)
 still gives an acceptable model
and there is clearly no fine tuning in these models.
The effect of a large number of extra degrees
of freedom $g_{ex}\sim 10^3$  at the condensation scale
is to
drop the energy density from
$\Omp(\Lm_c)=0.07$ with $g_{ex}=64$ to $ \Omp(\Lm_c)=0.01
$ with $g_{ex}=10^3$ and the numerical solution, in this case,
 gives $\wpo=-0.82$ at
present time  still within the observational limits. In fact,
there is no upper limit for $g_{ex}$ from the cosmological
evolution constrains for $\phi$  because
the upper value for $\wpo$ is given by its tracker value
which for $n=2/3$ is $w_{tr}=-0.75 < -2/3$ smaller than
the upper limit given by SN1a and CMBR data.
In figure \ref{fig1} we show
the evolution of $\Omp,\wp$ for the minimal number
of $g_{ex}=64$ ($\Ompi=0.07$)
and for an extreme case with $g_{ex} \sim 10^9$ ($\Ompi=10^{-10}$)
and in both cases we get an acceptable model.

In table  \ref{tab1} and \ref{tab2} we summarize
the relevant cosmological quantities. In table
\ref{tab1} we give  the values of $n, b_o$, the
degrees of freedom of $Q$ with ($g_{Q s}$) and without supersymmetry ($g_{Q nsusy}$),
the condensation scale  $\Lm_c$. Notice that all models
have same $b_o, \Lm_c$ but $n$ differs slightly. Model I
is the minimal model, in the sense that it has the smallest
number of degrees of freedom.

In table \ref{tab2} we give the values of the initial
energy density $\Om_Q(\Lm_{gut})$, the energy density
at NS (for $g_{ex}=0)$, the number of extra degrees of freedom
needed to have $\Om_Q(NS)=0.1$ or 0.2, the value of $\Omp(\Lm_c)$ with
$g_{ex}=0$,  the value of $N_{Tot}$ (the e-folds from
$\Lm_c$ to present day), the values of $\wpo$ and
 $w_{eff} $ calculated numerically and
the value obtain analytically from eq.(\ref{wo})
 gives a good approximation to
 the numerical one. The energy density at $\Lm_c$
with the condition $\Om_Q(NS) \leq 0.1$ gives
 $\Om_{\Lm_c}=0.07$ for all three models while for $\Om_Q(NS) \leq 0.2$ gives
 $\Om_{\Lm_c}=0.15$ for models II and III, respectively.
Since the number of $g_{ex}$ for
 models II and III is quite large (larger than MSSM) we
consider them less "natural" then the minimal Model I.

\begin{table}
\begin{center}
 \begin{tabular}{|c|c|c|c|c|c|c|c|}\hline
   Num & $N_c$ & $N_f$ & $\nu$ & $n$ & $16\pi^2b_o$ & $\Lm_c (GeV)$
   & $g_{Q s}$ \\ \hline \hline
   I & 3 & 6 & 1& 2/3 & 3 & $4.2\times 10^{-8}$ & 97.5\\
   II & 6 & 15 & 3& 2/3 & 3 & $4.2\times 10^{-8}$ & 468.5\\
   III & 7 & 18 & 4& 6/11 & 3 & $4.2\times 10^{-8}$ & 652.5  \\
     \hline
 \end{tabular}
 \end{center}
\caption{\small{We show the matter content for the three different models
and we give the number of degrees of freedom for the susy and
non susy $Q$ group in the last two columns, respectively.
Notice that the condensation scale
and $b_o$ is the same for all models.}}
\la{tab1}\end{table}
\begin{table}
\begin{center}
\begin{tabular}{|c|c|c|c|c|c|c|c|c|c|}\hline
  Num & $\Om_Q(\Lm_{gut})$ & $\Om_Q(NS) $ &
   $\Omp(\Lm_{c}) $ &$g_{ex 1 }$
   & $g_{ex 2}$ &$\wpo$  & $\weff$ & $\wpo(Th)$ & $N_{Tot}$
    \\ \hline \hline
  I & 0.30 & 0.13 & 0.09 &  64 & 0& -0.90 & -0.93 & -0.82 & 12.9 \\
  II & 0.67 & 0.42 & 0.33   & 718 &287& -0.90 & -0.93 & -0.82 & 12.9 \\
  III & 0.74 & 0.50 & 0.41   & 986 & 433&-0.93 & -0.95 & -0.87 & 11.1 \\
     \hline
\end{tabular}
 \end{center}
\caption{\small{The first column  gives the  model number.
In columns 2-4 we give the energy density at different scales
assuming no extra degrees  of freedom (i.e. $g_{ex}=0$). In column 5 and 6 we
show the  necessary number of
$g_{ex}$ to have $\Om_Q(NS) \leq 0.1, 0.2$, respectively. We show in
columns 7, 8 the present day value of $\wp$ calculated numerically
and in column 9 the theoretically obtained from eq.(\ref{wo}). Finally, we give in the last
column the number of e-folds of expansion from $\Lm_c$ to present day. }}
\la{tab2}\end{table}

\section{Summary and Conclusions}

We have shown that an unification scheme, where
all coupling constants are unified, as predicted by string
theory, leads to an acceptable  cosmological constant
parameterized in terms of the condensates of
a non-abelian gauge group. These fields play the role
of quintessence.

  Above the unification scale we
have all fields  in thermal equilibrium and the number
of degrees of freedom for the SM and $Q$ model determines
the initial conditions for each group. Below $\Lm_{gut}$
the $Q$ group decouples, since it interacts with the SM
only through gravity. For
temperatures above the condensation scale of the $Q$ group
its fields are relativistic and red shift as radiation.
The entropy of each systems is independently
conserved and we can therefore determine the energy
density at NS and at $\Lm_c$. The models we have
obtain have a condensation scale below NS and in order
not to spoil the NS predictions the energy density
must be $\Om_Q(NS) <0.1-0.2$.

Without considering the contribution from
the susy breaking sector, all models have $\Om_Q > 0.1$ with
the smallest contribution from Model I ($N_c=3,N_f=6$ minimal model)
giving
$\Om_Q =0.13$ in agreement with the conservative bound $\Om_Q|_{NS} <0.2$
NS bound and slightly large then $\Om_Q|_{NS} <0.1$. If susy
is transmitted via gravity we require extra structure to
agree with the strongest NS (the gauge group responsible for
susy breaking is not in thermal equilibrium with the SM
below $\Lm_{gut}$) but if susy is gauge mediated than the NS bound
is alleviated
 since one has extra degrees of freedom $g_{ex}$ in thermal equilibrium
 with the SM.  The cosmological evolution of quintessence is not sensitive
 to the number of the extra degrees of freedom. There is a minimum
 number required  from NS bounds but  there is no upper limit.

At the condensation scale the $Q$ fields are no longer free and
they condense. We use Affleck's potential to parameterize the
condensates and we study the cosmological evolution with the
initial condition determined in terms of $N_c, N_f$ only. Gauge
unification determines the values of $N_c, N_f$ and there are no
models with $2\times 10^{-2}
GeV < \Lm_c < 6 \times 10^3 GeV$ or  $2<n<4.27$. Since
$\wpo<-2/3$ requires $n<2.74$ all models must have
$\Lm_c > 2\times 10^{-2} GeV$ or $n>2$. The three
acceptable models have a potential of the form $V \sim \phi^{-n}$
with $6/11\leq n \leq 2/3$. The value of $n$ and the energy
density at $\Lm_c$ determines the present day value of $\wpo$. We
show that the models  have $\Ompo=0.7, \wpo=-0.90$ with a Hubble
parameter $h_o=0.7$ and the value of $n$ and $\wpo$ are in
accordance with constrains from recent CMBR analysis, i.e. $n<1$
and $\wpo=-0.82^{+.14}_{-.11}$  \ci{neww,wette}. We  also show that
the tracker
solution to inverse power potential is not specific enough (in Model I
$w_{tr}=-0.75$) and does not give a good approximation for models
with $n<2$, which are the cosmologically favored.

We would like to stress out that there are no free parameters,
not even the $Q$ initial energy density at unification nor at the
condensation scale. The models are well motivated  from a
particle physics point of view, they involve a late time
phase transition, and they agree with present day
observations.

This work was supported in part by CONACYT project 32415-E and
DGAPA, UNAM project IN-110200.

\thebibliography{}

\footnotesize{

\bib{CMBR} {P. de Bernardis {\it et al}. Nature, (London) 404, (2000)
955, S. Hannany {\it et al}.,Astrophys.J.545 (2000) L1-L4}

\bib{SN1a} {A.G. Riess {\it et al.}, Astron. J. 116 (1998) 1009; S.
Perlmutter {\it et al}, ApJ 517 (1999) 565; P.M. Garnavich {\it et
al}, Ap.J 509 (1998) 74.}

\bib{structure}{ G. Efstathiou, S. Maddox and W. Sutherland,
 Nature 348 (1990) 705.
  J. Primack and A. Klypin, Nucl. Phys. Proc. Suppl. 51 B, (1996),
30}

\bib{w}{S. Perlmutter, M. Turner and  M. J. White,
Phys.Rev.Lett.83:670-673, 1999; T. Saini, S. Raychaudhury, V. Sahni
and  A.A. Starobinsky, Phys.Rev.Lett.85:1162-1165,2000 }

\bib{neww} Carlo Baccigalupi, Amedeo Balbi, Sabino Matarrese,
Francesca Perrotta, Nicola Vittorio, astro-ph/0109097

\bib{wette} Michael Doran, Matthew J. Lilley, Jan Schwindt,
 Christof Wetterich, astro-ph/0012139; Michael Doran, Matthew Lilley,
  Christof Wetterich  astro-ph/0105457

\bib{wcop}  P.S. Corasaniti, E.J. Copeland, astro-ph/0107378

\bib{tracker} I. Zlatev, L. Wang and P.J. Steinhardt, Phys. Rev.
Lett.82 (1999) 8960;  Phys. Rev. D59 (1999)123504

\bib{1/q} {P.J.E. Peebles and B. Ratra, ApJ 325 (1988) L17;
Phys. Rev. D37 (1988) 3406}

\bib{2/q}{J.P. Uzan, Phys.Rev.D59:123510,1999}

\bib{bine}{P. Binetruy, Phys.Rev. D60 (1999) 063502, Int. J.Theor.
Phys.39 (2000) 1859}

\bib{mas} {A. Masiero, M. Pietroni and F. Rosati, Phys.
Rev. D61 (2000) 023509}

\bib{chris1}{A. de la Macorra and C. Stephan-Otto, Phys.Rev.Lett.87:271301,2001 }

\bib{generic}  A.R. Liddle and R.J. Scherrer, Phys.Rev.
D59,  (1999)023509

\bib{mio.scalar}{A. de la Macorra and G. Piccinelli, Phys.
Rev.D61 (2000) 123503}

\bib{liddle}E.J. Copeland, A. Liddle and D. Wands, Phys. Rev. D57 (1998) 4686

\bib{brax}  P. Brax, J. Martin, Phys.Rev.D61:103502,2000

\bib{r=m}{E.Kolb and M.S Turner,The Early Universe, Edit. Addison Wesley 1990}

\bib{NS} {K. Freese, F.C. Adams, J.A. Frieman and E. Mottola, Nucl. Phys. B 287
(1987) 797; M. Birkel and S. Sarkar, Astropart. Phys. 6 (1997)
197.}

\bib{NSC} {C. Wetterich, Nucl. Phys. B302 (1988) 302, R.H. Cyburt, B.D. Fields,   K. A. Olive, Astropart.Phys.17:87-100,2002  }

\bib{Affleck}{I. Affleck, M. Dine and N. Seiberg, Nucl. Phys.B256
(1985) 557}

\bib{shifman} M.A. Shifman and A.I. Vainshtein, Nucl. Phys. B277
(1986) 649

\bib{ax.asy}{C.P. Burgess, A. de la Macorra, I. Maksymyk and F. Quevedo
Phys.Lett.B410 (1997) 181}

\bib{ax.coinci}{A. de la Macorra, Int.J.Mod.Phys.D9 (2000) 661 }

\bib{chris2}{ A. de la Macorra and C. Stephan-Otto, Phys.Rev.D65:083520,2002 }

\bib{unif}{U. Amaldi, W. de Boer and H. Furstenau, Phys. Lett.B260
(1991) 447, P.Langacker and M. Luo, Phys. Rev.D44 (1991) 817}
}

\bib{duality}{K. Intriligator and  N. Seiberg, Nucl.Phys.Proc.Suppl.45BC:1-28,1996

\bib{gaugino} M. Dine, R. Rhom, N. Seiberg and E. Witten, Phys.
Lett. B156 (1985) 55, G. Veneziano and S. Yankielowicz, Phys.
Lett. B113 (1984)231; S. Ferrara, L. Girardello and H.P. Nilles, Phys. Lett.
B125 (1983) 457, D. Amati, K. Konishi, Y. Meurice, G. Rossi
and G. Veneziano, Phys. Rep. 162 (1988) 169

\bib{sus.grav}T.R. Taylor, Phys.Lett.B252 (1990) 59;H. P. Nilles
Int.J.Mod.Phys.A5 (1990) 4199; B. de Carlos,
  J.A. Casas, C. Munoz,  Nucl.Phys.B399 (1993) 623;
  M. Cvetic, A. Font, Luis E. Ibanez, D. Lust), F. Quevedo
 Nucl.Phys.B361 (1991) 194, Phys.Lett.B245 (1990) 401;
 A. de la Macorra and G.G. Ross, Phys. Lett. B
Nucl.Phys.B404 (1993) 321, Nucl.Phys.B443 (1995)127-154

\bib{sus.gauge}{G.F. Giudice, R. Rattazzi,
Phys.Rept.322:(1999) 419-499 and ref. therein}

\end{document}